\documentclass[aip,rsi,superscriptaddress,amsmath,amssymb,reprint,longbibliography]{revtex4-2}
\usepackage{graphicx}
\usepackage{lettrine}
\usepackage{adjustbox}
\usepackage{multirow}
\usepackage{xcolor}
\usepackage[colorlinks=true,
            linkcolor=blue,
            citecolor=blue,
            urlcolor=blue]{hyperref}

\begin{document}
\title{Scaling laws for contamination traversal times during vacuum breaks in superconducting accelerators}

\author{Yinghe Qi}
\affiliation{National High Magnetic Field Laboratory, Tallahassee, FL 32310, USA}
\affiliation{Department of Mechanical and Aerospace Engineering, FAMU-FSU College of Engineering, Florida State University, Tallahassee, FL 32310, USA}

\author{Qutadah Rababah}
\affiliation{National High Magnetic Field Laboratory, Tallahassee, FL 32310, USA}
\affiliation{Department of Mechanical and Aerospace Engineering, FAMU-FSU College of Engineering, Florida State University, Tallahassee, FL 32310, USA}

\author{Wei Guo}
\email[Email: ]{wguo@famu.fsu.eng.edu}
\affiliation{National High Magnetic Field Laboratory, Tallahassee, FL 32310, USA}
\affiliation{Department of Mechanical and Aerospace Engineering, FAMU-FSU College of Engineering, Florida State University, Tallahassee, FL 32310, USA}


\begin{abstract}
A sudden loss of vacuum in a superconducting radio-frequency (SRF) accelerator drives warm gas into the evacuated beamline, where it propagates through interconnected SRF cavities cooled by liquid helium, causing surface contamination, intense heat deposition, and dangerous pressure buildup. Yet predicting how such a failure propagates through realistic multicell cavities has remained difficult because compressible gas dynamics, condensation and cryogenic heat transfer are strongly coupled. Here we develop and experimentally validate a multidimensional computational framework that resolves these processes, and apply it to a full-scale nine-cell SRF cavity. We find that the repeated cell geometry produces staircase-like gas-front propagation, as impingement, recirculation and localized cryopumping strongly retard the bulk flow within each cell. Across a broad range of breach sizes, the cavity traversal time follows a robust power-law scaling with leakage area, providing quantitative guidance for the response time of beamline isolation and protection systems. The simulations also reveal a geometry-confined Mach disk whose position and size track the advancing gas front, suggesting that its acoustic emission could potentially enable rapid, non-invasive detection and localization of vacuum failures. These findings pave the way for a new generation of predictive diagnostics and protection strategies to safeguard large-scale superconducting accelerators.
\end{abstract}

\maketitle

\section*{Introduction}\label{Sec_Intro}
\noindent Superconducting radio-frequency (SRF) cavities form the backbone of many modern high-energy particle accelerators, where reliable operation requires ultraclean beamlines maintained under ultra-high vacuum and at cryogenic temperatures \cite{padamsee2015design,aune2000superconducting,peterson2015lcls}. As shown in Fig.~\ref{Fig1}a, a typical SRF cryomodule contains a string of multicell cavities connected through a common beam vacuum and cooled by surrounding liquid helium (LHe), with many cryomodules linked in series to form an extended accelerator beamline. In facilities such as the European XFEL and LCLS-II, these beamlines extend over kilometer scales \cite{EuropeanXFEL,LCLSII}, making a sudden loss of beamline vacuum (SLV) a particularly serious failure scenario. Following a breach, room-temperature gas rushes into the evacuated beamline and rapidly condenses on the cryogenic cavity inner surfaces. The resulting heat load drives intense helium boil-off and pressure buildup, while the advancing gas can transport frost, particulates, and other contaminants deep into the cavity string \cite{wisemanLossCavityVacuum1994,adyLeakPropagationDynamics2014,boeckmannExperimentalTestsFault2008,seidelFailureAnalysisBeam2002,wiseman1991cebaf}. Such an event threatens both cryogenic safety and the stringent cleanliness required for SRF operation, potentially degrading cavity performance and causing prolonged accelerator downtime. Limiting its extent requires rapid detection, isolation, and pressure relief \cite{seidelFailureAnalysisBeam2002,zapfe2007vacuum,adyLeakPropagationDynamics2014,dalesandroExperimentTransientEffects2012,lehmann1978safety}. The effectiveness of these protection measures ultimately depends on a fundamental time scale: how rapidly does the invading contamination front propagate through a cryogenic accelerator beamline?

Predicting this propagation time turns out to be a challenging task because the underlying processes are strongly coupled. Cryocondensation continuously removes mass, momentum, and energy from the invading gas, causing the cryogenic wall to act as a distributed cryopump \cite{dawsonCryopumping1965,welchCapturePumpingTechnology2001}. Meanwhile, the growing frost layer and associated heat transfer alter the wall temperature and local condensation rate, creating a dynamic feedback between gas flow, deposition, and the surrounding LHe \cite{baoHeatMassTransfer2020,garceauHeatMassTransfer2021,van_sciver_helium_2012}. This coupling is further complicated by the repeated contractions and expansions of multicell SRF cavities, which produce jets, recirculation, and shocks. The contamination traversal time is thus governed not simply by gas expansion, but by the complex, transient interplay of all these physical processes.



Early vacuum-loss experiments at CEBAF, the European XFEL, Fermilab, and other facilities established that gas propagation during an SLV can be dramatically slower than expansion into a warm vacuum~\cite{wisemanLossCavityVacuum1994,adyLeakPropagationDynamics2014,boeckmannExperimentalTestsFault2008,dalesandroExperimentTransientEffects2012,dalesandroResultsSuddenLoss2014}. These experiments provided valuable insight under realistic accelerator conditions, but their scale and complexity make systematic variation of the governing parameters difficult. This motivated our controlled laboratory studies using uniform LHe-cooled tubes, where the leak conditions and helium state could be varied systematically. These experiments quantified the strong deceleration of the gas front and enabled one-dimensional models that successfully reproduced measured gas-front trajectories, determined the characteristic heat load, and predicted the freeze range~\cite{dhuleySuddenVacuumLoss2014,dhuleyPropagationNitrogenGas2016a,dhuleyPropagationNitrogenGas2016,dhuleyGasPropagationLiquid2016,garceauHeatMassTransfer2019,garceauEffectMassFlow2020,garceauHeatMassTransfer2021,baoHeatMassTransfer2020,bao2024freeze}. The uniform-pipe simplification that made these studies systematic, however, also removed the repeated expansions and restrictive irises of actual SRF cavities, where multidimensional flow can substantially alter gas transport and deposition. A predictive, cavity-resolved framework connecting leak conditions to contamination traversal time in realistic SRF beamlines has therefore remained lacking.

Here we bridge this gap by developing a cavity-resolved numerical framework, validating it against controlled vacuum-break measurements in both uniform and non-uniform tube systems, and applying it to a full-scale nine-cell TESLA-type SRF cavity. The simulations reveal how the repeated cavity geometry reshapes the advancing flow through strong impingement and recirculation, producing a distinct cell-by-cell progression of the macroscopic contamination front. By varying the leakage area, we identify a robust scaling law for the cavity traversal time that links breach severity to the response time window available for beamline isolation. The simulations further reveal an intriguing Mach disk confined near the cavity iris whose position and size evolve with the advancing contamination front. This coupling between a localized shock structure and the global propagation dynamics suggests that its acoustic signature could potentially enable rapid, non-invasive tracking of vacuum-break events. Together, these results connect cavity-scale SLV dynamics to practical strategies for protecting and monitoring superconducting accelerator beamlines.

\section*{Results}\label{SecII}
\noindent\textbf{Multidimensional modeling of vacuum-break propagation}\\
\noindent To describe SLV propagation beyond the uniform-tube approximation, we extended our previous one-dimensional model~\cite{baoHeatMassTransfer2020,garceauHeatMassTransfer2021,bao2024freeze} to a two-dimensional axisymmetric framework that resolves both axial and radial momentum transport. This allows flow separation, recirculation, jetting, and shock formation to emerge naturally in non-uniform beamline geometries. We adopt the Reynolds-averaged conservation equations for mass, momentum, and total energy to describe the transient evolution of the compressible gas flow~\cite{Pope2000,FerzigerPeric2002}:
\begin{flalign}
&\frac{\partial \rho_\text{g}}{\partial t}+\nabla\cdot(\rho_\text{g}\mathbf{u})=S_m,
&&\label{eq:GE_mass}\\[0.1\baselineskip]
&\frac{\partial(\rho_\text{g}\mathbf{u})}{\partial t}+\nabla\cdot(\rho_\text{g}\mathbf{u}\mathbf{u})=-\nabla P_g+\nabla\cdot\boldsymbol{\tau}_{\mathrm{eff}}+\mathbf{S}_u,
&&\label{eq:GE_momentum}\\[0.1\baselineskip]
&\begin{aligned}
\frac{\partial(\rho_\text{g}E)}{\partial t}+\nabla\cdot\Bigl[\mathbf{u}(\rho_\text{g}E+
&P_g)\Bigr]=\nabla\cdot\left(k_{\mathrm{eff}}\nabla T_g\right)\\
&\quad+\nabla\cdot\left(\boldsymbol{\tau}_{\mathrm{eff}}\cdot\mathbf{u}\right)+S_h .
\end{aligned}
&&\label{eq:GE_energy}
\end{flalign}
Here $\rho_g$, $\mathbf{u}$, and $T_g$ denote the gas density, velocity, and temperature, respectively. The total specific energy is $E=E_i+\frac{1}{2}|\mathbf{u}|^2$, where the specific internal energy $E_i$ and gas pressure $P_g$ can be obtained from tabulated property data as functions of $\rho_{\mathrm{g}}$ and $T_g$.

The effective stress tensor $\boldsymbol{\tau}_{\mathrm{eff}}$ and effective thermal conductivity $k_{\mathrm{eff}}$ appearing in Eqs.~(\ref{eq:GE_momentum}) and (\ref{eq:GE_energy}) are defined as $\boldsymbol{\tau}_{\mathrm{eff}}=\mu_{\mathrm{eff}}\left[\nabla\mathbf{u}+(\nabla\mathbf{u})^{T}-\frac{2}{3}(\nabla\cdot\mathbf{u})\mathbf{I}\right]$ and $k_{\mathrm{eff}}=k_{\mathrm{g}}+c_p\mu_t/Pr_t$, respectively~\cite{Pope2000,FerzigerPeric2002}. Here $\mu_{\mathrm{eff}}=\mu+\mu_t$ is the effective dynamic viscosity, with $\mu$ and $\mu_t$ denoting the gas molecular and turbulent viscosities, respectively; $k_{\mathrm{g}}$ is the molecular thermal conductivity of the gas, $c_p$ is its specific heat, and $Pr_t$ is the turbulent Prandtl number. The turbulent viscosity $\mu_t$, which enters both $\mu_{\mathrm{eff}}$ and $k_{\mathrm{eff}}$, is evaluated using the standard $k$--$\varepsilon$ model~\cite{Launder1972,Launder1974} as $\mu_t=C_\mu\rho_{\mathrm{g}}k^2/\varepsilon$, where $k$ is the turbulent kinetic energy, $\varepsilon$ is its dissipation rate, and $C_\mu=0.09$ is the standard model constant~\cite{Launder1974}. The transport equations for $k$ and $\varepsilon$ are solved concurrently with the mean-flow equations; their full forms together with the turbulence-production term are provided in Methods.

\renewcommand{\figurename}{\textbf{Fig.}}
\begin{figure*}[t]
\centering
\includegraphics[width=1\linewidth]{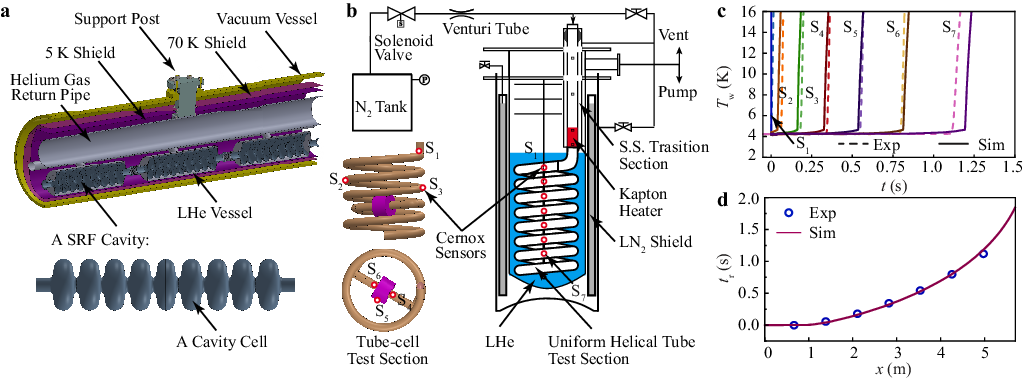}
\caption{\textbf{Experimental apparatus and uniform-tube validation of the vacuum-break model.}
\textbf{a} Schematic of a TESLA-type cryomodule showing the string of multicell superconducting radio-frequency (SRF) cavities connected through a common beam vacuum.
\textbf{b} Schematic of our vacuum-break experimental apparatus. The uniform helical tube is shown installed in the LHe cryostat, while the alternative tube-cell test section is illustrated separately at left, together with the surface thermometers used to track gas-front propagation indicated along the tube.
\textbf{c} Measured and simulated transient wall-temperature responses at the sensor locations along the uniform tube following a 100-kPa vacuum breach.
\textbf{d} Corresponding gas-front arrival time $t_r$ as a function of streamwise distance $x$, with the local arrival defined by $T_{\mathrm{w}}=4.7$~K.}
\label{Fig1}
\end{figure*}

The remaining source terms in Eqs.~(\ref{eq:GE_mass})--(\ref{eq:GE_energy}) describe the coupling of cryocondensation to the gas dynamics. An wall-surface condensation flux $\dot{m}_{\mathrm{c}}$ produces a volumetric mass sink $S_m=-\dot{m}_{\mathrm{c}}/\delta_{\mathrm{grid}}$ in the first layer of fluid cells adjacent to the cryogenic wall, where $\delta_{\mathrm{grid}}$ is the wall-normal thickness of the adjacent cell. The associated momentum and energy sinks are $\mathbf{S}_u=S_m\mathbf{u}$ and $S_h=S_mE$, respectively. Cryocondensation is treated consistently in the turbulence equations through the corresponding sinks $S_k=S_mk$ and $S_\varepsilon=S_m\varepsilon$. Thus, deposition removes mass together with the momentum, energy, and turbulent quantities carried by that mass.

For closure, the wall-surface condensation flux is determined using the Hertz–Knudsen relation~\cite{Schrage1953}:
\begin{equation}
\dot{m}_{\mathrm{c}}=\sqrt{\frac{M_{\mathrm{g}}}{2\pi R}}\left[\sigma_{\mathrm{c}}\frac{P_{\mathrm{g}}}{\sqrt{T_{\mathrm{g}}}}-\sigma_{\mathrm{e}}\frac{P_{\mathrm{sat}}(T_{\mathrm{s}})}{\sqrt{T_{\mathrm{s}}}}\right],
\label{eq:mc}
\end{equation}
where $M_{\mathrm{g}}$ is the gas molar mass, $R$ is the universal gas constant, $T_{\mathrm{s}}$ is the instantaneous frost-surface temperature, and $P_{\mathrm{sat}}(T_{\mathrm{s}})$ is the corresponding saturation vapor pressure. The coefficients $\sigma_{\mathrm{c}}$ and $\sigma_{\mathrm{e}}$ are the condensation and evaporation coefficients, respectively. While $\sigma_{\mathrm{e}}$ is typically close to unity~\cite{garceauHeatMassTransfer2021}, $\sigma_{\mathrm{c}}$ depends on the local degree of supercooling, which we model as $\sigma_{\mathrm{c}}=SD^h$. Here, $SD=[T_{\mathrm{sat}}(P_{\mathrm{g}})-T_{\mathrm{s}}]/T_{\mathrm{sat}}(P_{\mathrm{g}})$ is the degree of supercooling, and $h$ is an empirical exponent to be determined from the uniform-tube validation described below. Throughout our simulations, the frost surface remains supercooled ($SD>0$); as it warms, $SD$ decreases and cryocondensation weakens.

The frost-surface temperature $T_{\mathrm{s}}$ in Eq.~(\ref{eq:mc}) is not prescribed, but evolves dynamically through the coupled thermal response of the growing frost layer, the solid wall, and the surrounding LHe bath. The frost thickness increases according to $\partial\delta_{\mathrm{SN}}/\partial t=\dot{m}_{\mathrm{c}}/\rho_{\mathrm{SN}}$, where $\rho_{\mathrm{SN}}$ is the density of frost layer~\cite{baoHeatMassTransfer2020,garceauHeatMassTransfer2021}. Meanwhile, $T_{\mathrm{s}}$ and the wall temperature $T_{\mathrm{w}}$ are obtained simultaneously from transient energy balances that account for energy deposited by the condensing gas, gas-to-frost heat transfer, conduction through the frost layer, and heat rejection from the outer wall to the LHe bath. This thermal coupling closes the feedback between cryocondensation and wall heating. The full thermal model and its numerical implementation are given in Methods.

\vspace{12pt}
\noindent\textbf{Experimental validation of the model}\\
\noindent
We validated the model in two stages, beginning with the extensively characterized uniform-tube configuration, where the streamwise propagation provides a controlled setting for constraining the cryocondensation model and determining the optimal parameter $h$. We then tested the model with the fixed $h$ against the helix-cavity configuration to assess its ability to capture geometry-induced multidimensional transport.

For the first stage, we used the uniform helical copper tube, with an inner diameter of $D_i=25.4$~mm and a length of 5.75~m, previously characterized over a broad range of vacuum-break conditions~\cite{garceauDesignTestingLiquid2019,garceauHeatMassTransfer2019,garceauEffectMassFlow2020,garceauHeatMassTransfer2021}. The experimental apparatus (Fig.~\ref{Fig1}b) consists of a 230-L nitrogen reservoir initially at 100~kPa, a fast-acting solenoid valve, and an inline venturi that regulates the mass flow. The incoming gas passes through a temperature-controlled stainless-steel transition section maintained above 77~K before entering the copper test section immersed in a 4.2-K LHe bath. Although operational SRF cavities are generally cooled by superfluid helium (He II), our previous uniform-tube measurements showed only minor differences in gas-front propagation between normal LHe and He II~\cite{garceauHeatMassTransfer2019,garceauHeatMassTransfer2021}.

For a 100-kPa vacuum breach, the calculated transient wall-temperature responses closely reproduce the measurements at the sensor locations along the uniform tube (Fig.~\ref{Fig1}c). We define the local arrival of the gas front by the wall-temperature threshold $T_{\mathrm{w}}=4.7$~K. The corresponding arrival times increase strongly and nonlinearly with streamwise distance as continuous cryocondensation progressively removes mass from the advancing gas and decelerates the front. The simulation captures this behavior closely (Fig.~\ref{Fig1}d). An optimal value of $h=1.3$ was obtained by minimizing the overall discrepancy between the measured and calculated arrival times.

\begin{figure*}[t]
\centering
\includegraphics[width=1\linewidth]{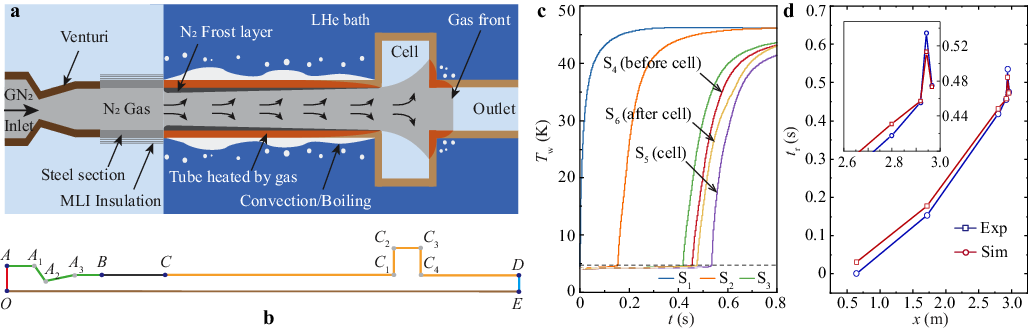}
\caption{\textbf{Multidimensional model and tube-cell validation.}
\textbf{a} Schematic of the coupled gas-flow, cryocondensation, frost-growth, wall-heating, and LHe heat-transfer processes included in the model.
\textbf{b} Two-dimensional axisymmetric computational domain for the tube-cell test section. Detailed dimensions and boundary conditions are provided in Supplementary Table~S1.
\textbf{c} Simulated transient wall-temperature responses at the experimental sensor locations (see Fig.~\ref{Fig1}b).
\textbf{d} Measured and simulated gas-front arrival times as a function of sensor position $x$, with the local arrival defined by $T_{\mathrm{w}}=4.7$~K. The inset enlarges the region around the cell to resolve the arrival times at the closely spaced sensors $S_4$, $S_5$, and $S_6$.
}
\label{Fig2}
\end{figure*}

For the second stage, the uniform helical tube test section was replaced by a custom tube-cell section containing a cylindrical copper cell between the third and fourth turns (Fig.~\ref{Fig1}b). The tube inner diameter remains $D_i=25.4$~mm, while the cell inner diameter and length are set to $3D_i$ and $1.8D_i$, respectively, to mimic the characteristic geometric proportions of a typical SRF cavity cell~\cite{padamsee2015design}. The remaining apparatus and the 100-kPa venting condition were unchanged. Six surface thermometers were mounted along the test section, with $S_4$ located immediately upstream of the cell, $S_5$ on the cell lateral wall, and $S_6$ immediately downstream of the cell exit (Fig.~\ref{Fig1}b), allowing axial penetration through the cell and radial spreading toward the wall to be detected experimentally. The corresponding two-dimensional computational domain and thermal boundary conditions are shown in Fig.~\ref{Fig2}a and b.

\renewcommand{\figurename}{\textbf{Fig.}}

The cell produces a clear departure from the monotonic propagation observed in the uniform tube. Experimentally, the gas front reaches the downstream sensor $S_6$ before the lateral-wall sensor $S_5$ (see Fig.~\ref{Fig2}), even though $S_6$ lies farther from the cell inlet. This arrival-time inversion indicates that the incoming gas penetrates rapidly through the central region of the expanded cell and reaches the downstream exit before sufficient radial transport occurs for substantial condensation on the cell lateral wall. The distinct time scales of axial penetration and radial filling provide direct experimental evidence of multidimensional SLV propagation.

Using the same $h=1.3$ determined from the uniform-tube data, our tube-cell simulation reproduces the experimentally observed behavior nicely. The calculated wall-temperature traces at the sensor locations recover the $S_4\rightarrow S_6\rightarrow S_5$ arrival sequence (Fig.~\ref{Fig2}c), while the arrival times extracted from these traces closely match the measured values at the corresponding positions (Fig.~\ref{Fig2}d). This agreement demonstrates that our model captures the coupled axial and radial transport responsible for the observed arrival-time inversion, without any geometry-specific adjustment.

\renewcommand{\figurename}{\textbf{Fig.}}
\begin{figure*}[t]
\centering
\includegraphics[width=1\linewidth]{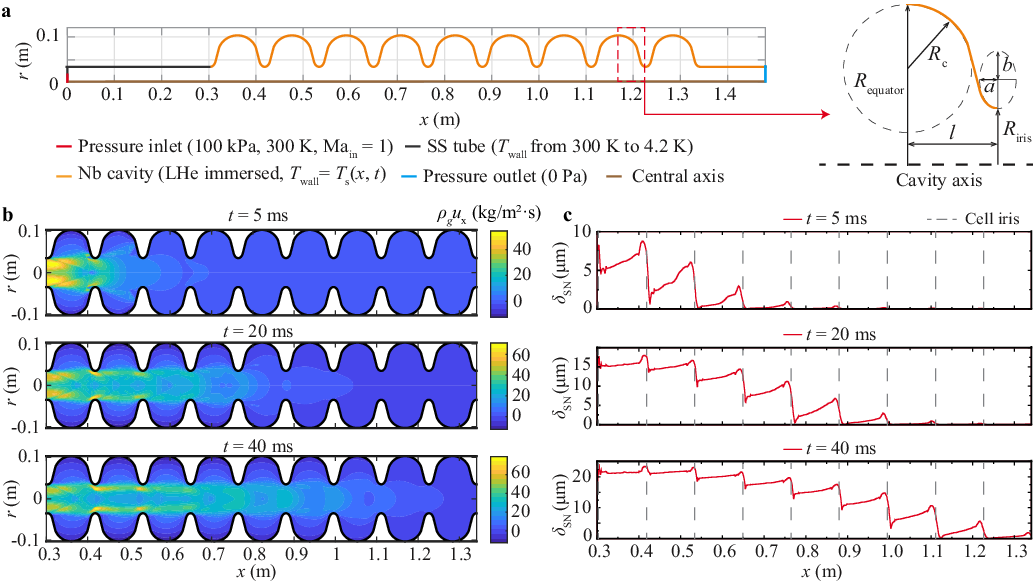}
\caption{\textbf{Transient gas-flow and frost-deposition dynamics in a nine-cell TESLA cavity.}
\textbf{a} Computational domain and cavity geometry. The full nine-cell cavity and representative half-cell are shown with the standard TESLA mid-cell dimensions~\cite{aune2000superconducting}: half-cell length $l=57.7$~mm, equator radius $R_{\mathrm{equator}}=103.3$~mm, iris radius $R_{\mathrm{iris}}=35$~mm, circular-arc radius $R_{\mathrm{c}}=42$~mm, and elliptical-arc semi-axes $a=12$~mm and $b=19$~mm. The segmented thermal boundaries and choked inlet condition are also indicated.
\textbf{b} Transient axial mass-flux ($\rho_g u_x$) fields at 5, 20, and 40~ms for a vacuum breach with leakage-area ratio $L_{\mathrm{A}}=0.2$. 
\textbf{c} Spatiotemporal evolution of the solid-nitrogen frost thickness $\delta_{\mathrm{SN}}$ along the cavity wall at 5, 20, and 40~ms.}
\label{Fig3}
\end{figure*}

\vspace{12pt}
\noindent\textbf{Multidimensional propagation and traversal-time scaling in a realistic SRF cavity}\\
\noindent With the model validated, we next applied it to a full-scale nine-cell TESLA-type SRF cavity, adopting the standard niobium cavity geometry and dimensions shown in Fig.~\ref{Fig3}a~\cite{aune2000superconducting}. A vacuum breach was introduced at the inlet, with its severity characterized by the leakage-area ratio $L_{\mathrm{A}}=S_{\mathrm{rupture}}/S_{\mathrm{tube}}$, where $S_{\mathrm{rupture}}$ denotes the rupture area and $S_{\mathrm{tube}}=\pi D_i^2/4$. We model the vacuum breach as a choked inlet, prescribing $P_0=100$~kPa, $T_0=300$~K, and an inlet Mach number $\mathrm{Ma}_{\mathrm{in}}=|\mathbf{u}|/c_g=1$, where $c_g$ is the local speed of sound. The cavity is initially evacuated and cooled to 4.2~K, while the upstream steel section spans the thermal gradient from room temperature to the cryogenic cavity. We first examine a representative severe breach with $L_{\mathrm{A}}=0.2$ and subsequently vary $L_{\mathrm{A}}$ from 0.05 to 0.50 to determine how the propagation depends on breach size.

\renewcommand{\figurename}{\textbf{Fig.}}
\begin{figure*}[t]
\centering
\includegraphics[width=1\linewidth]{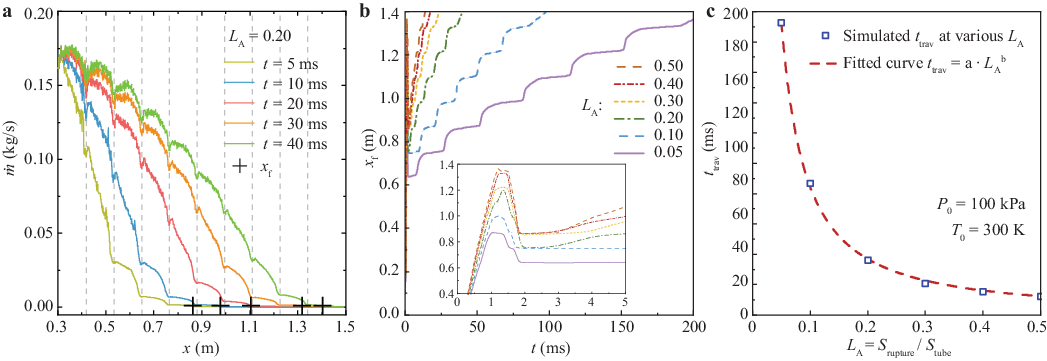}
\caption{\textbf{Macroscopic gas-front propagation and traversal-time scaling in a nine-cell TESLA cavity.}
\textbf{a} Instantaneous axial mass-flow-rate ($\dot{m}$) profiles at selected times for a vacuum breach with leakage-area ratio $L_{\mathrm{A}}=0.2$. Vertical grey dashed lines mark the cavity irises. The macroscopic gas-front position $x_{\mathrm{f}}$ is defined by the threshold $\dot{m}=0.001$~kg~s$^{-1}$ and indicated by $+$ symbols.
\textbf{b} Trajectories of $x_{\mathrm{f}}$ for leakage-area ratios $L_{\mathrm{A}}=0.05$--$0.50$. The inset enlarges the first 5~ms, revealing a rapidly propagating low-density precursor ahead of the bulk gas front. The subsequent trajectories exhibit a staircase-like progression associated with cell-by-cell propagation through the cavity.
\textbf{c} Cavity traversal time $t_{\mathrm{trav}}$, defined by the arrival of $x_{\mathrm{f}}$ at the exit of the ninth cell, as a function of $L_{\mathrm{A}}$. The dashed curve is a power-law fit $t_{\mathrm{trav}}=aL_{\mathrm{A}}^{b}$, with $a=5.4$~ms and $b=-1.19$.}
\label{Fig4}
\end{figure*}

For $L_{\mathrm{A}}=0.2$, the macroscopic gas front traverses the nine-cell cavity in approximately 40~ms, but the propagation is far from a simple axial advance. The axial mass-flux fields plotted in Fig.~\ref{Fig3}b reveal a repeated sequence of jetting, impingement, flow separation, and recirculation. After passing through each restrictive iris, the gas expands into the larger cell volume and impinges on the downstream wall near the junction of the elliptical and circular arcs. The flow then separates into two branches: an outer branch turns radially outward and recirculates toward the equator, producing regions of negative $\rho u_x$ (see corresponding velocity-vector fields in Supplementary Fig.~S1), whereas an inner branch turns toward the axis and converges into a strong forward jet approaching the next iris. This sequence repeats from cell to cell, demonstrating how the cavity geometry continuously redistributes axial momentum into radial motion and recirculation. The full temporal evolution of the axial mass-flux field for $L_{\mathrm{A}}=0.1$, 0.3, and 0.5 are shown in Supplementary Video~1.

The same flow structure strongly shapes the initial deposition of solid nitrogen. As shown in Fig.~\ref{Fig3}c, the frost thickness $\delta_{\mathrm{SN}}$ is highly non-uniform when the gas first enters a cell, with deposition concentrated on the downstream wall near the impingement region. The spatial correspondence among the frost thickness, mass-deposition rate, and wall-temperature response (Supplementary Fig.~S2) confirms that this localization is driven by the impinging flow. As the cell progressively fills, the deposition spreads over a larger surface area and the initial aerodynamic bias weakens. By 40~ms, the frost layer in the first several upstream cells has become substantially more uniform, reaching a characteristic thickness of about $20~\mu$m in the first three cells. Thus, the cavity geometry controls not only how the gas propagates but also where contamination and heat deposition are concentrated during the early stages of an SLV.

To quantify this complex propagation, we integrate the local axial mass flux across each transverse section of the axisymmetric domain,
$\dot{m}=\int \rho_g u_x\,2\pi r\,dr$, at axial intervals of $\Delta x=1$~mm. The resulting instantaneous mass-flow profiles for $L_{\mathrm{A}}=0.2$ are shown in Fig.~\ref{Fig4}a. Their oscillatory spatial structure reflects the repeated cavity geometry: although $\rho_g u_x$ becomes large near an iris because the flow converges there, the integrated mass flow $\dot{m}$ exhibits a local minimum because of the strongly reduced cross-sectional area. We then introduce the macroscopic gas-front position, $x_{\mathrm{f}}$, defined as the downstream-most axial location where $\dot{m}(x,t)$ reaches a threshold value of $10^{-3}$~kg~s$^{-1}$ (see Fig.~\ref{Fig4}a).

The resulting $x_{\mathrm{f}}(t)$ trajectories for $L_{\mathrm{A}}=0.05$--$0.50$ reveal two distinct propagation scales (Fig.~\ref{Fig4}b). During the first few milliseconds, a low-density precursor advances through the evacuated cavity much faster than the bulk gas, with an apparent propagation speed of order $10^3$~m~s$^{-1}$. Because this precursor carries very little mass, however, the subsequent motion of the bulk front governs the transport of contamination through the cavity. In contrast to the smooth deceleration observed in a uniform tube, the bulk front advances in a pronounced staircase-like manner. Each plateau corresponds to a period during which the incoming gas expands into a cell, impinges on the downstream wall, recirculates, and undergoes intense cryocondensation before sufficient gas reaches the next iris for the front to advance into the subsequent cell. The staircase trajectory therefore provides a macroscopic manifestation of the repeated multidimensional flow pattern resolved in Fig.~\ref{Fig3}b.

The duration of these cell-by-cell delays depends strongly on the breach size. We define the cavity traversal time $t_{\mathrm{trav}}$ as the time at which the bulk front reaches the exit of the ninth cell. As $L_{\mathrm{A}}$ decreases, $t_{\mathrm{trav}}$ rises rapidly and nonlinearly (Fig.~\ref{Fig4}c). Over the simulated range, the dependence is described remarkably well by
\[
t_{\mathrm{trav}}=(5.4~\mathrm{ms})\cdot L_{\mathrm{A}}^{-1.19},
\]
with $R^2>0.999$. This scaling directly connects the size of a vacuum breach to the time available before the macroscopic contamination front traverses the nine-cell cavity, providing a quantitative basis for evaluating the response times required for beamline isolation and machine-protection systems.

\begin{figure*}[t]
\centering
\includegraphics[width=1\linewidth]{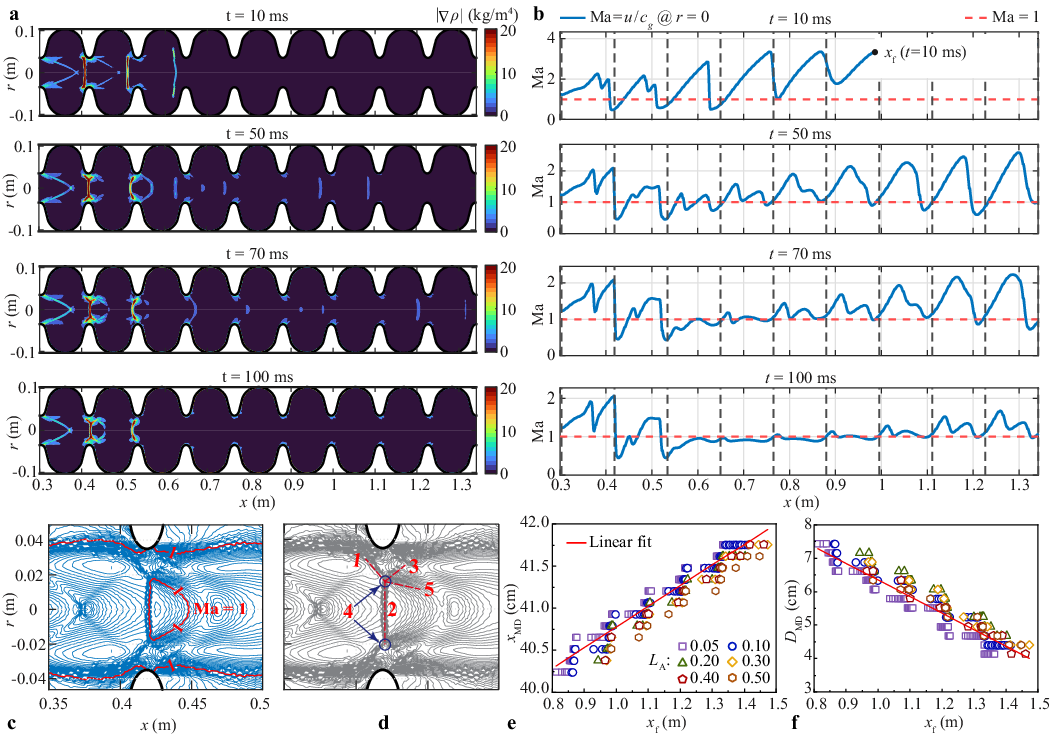}
\caption{\textbf{Supersonic flow dynamics and Mach-disk kinematics in a nine-cell TESLA cavity.}
\textbf{a} Simulated Schlieren fields ($|\nabla\rho|$) at selected times for a vacuum breach with $L_{\mathrm{A}}=0.2$. The displayed range is limited to 0--20~kg~m$^{-4}$ to enhance the visibility of shock structures. Persistent shocks remain near the first two irises, whereas downstream shocks propagate with the advancing gas front and weaken as the cavity fills.
\textbf{b} Corresponding Mach-number profiles along the cavity centerline ($r=0$). Sharp transitions from supersonic to subsonic flow coincide with the stationary and propagating shock structures identified in \textbf{a}.
\textbf{c} Magnified Mach-number contours showing the primary Mach disk near the first iris at $t=100$~ms.
\textbf{d} Structure of the confined shock system, showing the incident shock (1), normal Mach disk (2), reflected shock (3), triple point (4), and slip line (5).
\textbf{e} Axial position $x_{\mathrm{MD}}$ and \textbf{f}, diameter $D_{\mathrm{MD}}$ of the primary Mach disk as functions of the macroscopic gas-front position $x_{\mathrm{f}}$ for $L_{\mathrm{A}}=0.05$--$0.50$. Both quantities follow approximately linear trends with $x_{\mathrm{f}}$; solid lines denote linear fits.}
\label{Fig5}
\end{figure*}

\vspace{12pt}
\noindent\textbf{Shock dynamics and Mach-disk propagation}\\
\noindent
Besides the macroscopic gas transport, the simulations also reveal pronounced supersonic flow and shock formation within the cavity. The large pressure ratio between the incoming gas and the evacuated beamline drives strongly underexpanded flow through the restrictive irises. Simulated Schlieren fields based on $\nabla\rho_g$ and the corresponding centerline Mach-number profiles (Fig.~\ref{Fig5}a and b) reveal two distinct classes of shock structures. The first class consists of transient shocks that develop near the advancing gas front and migrate downstream as successive cells fill. At 10~ms, pronounced supersonic-to-subsonic transitions are evident across the first several irises; by 50~ms the dominant transient signatures have shifted toward the sixth and seventh irises, and by 70~ms toward the final cells. These downstream shocks progressively weaken as the local cavity pressure rises and the pressure ratio across the preceding irises decreases, eventually disappearing as the pressure field relaxes. The full evolution of the Schlieren field for $L_{\mathrm{A}}=0.1$, 0.3, and 0.5 are shown in Supplementary Video~2.

\renewcommand{\figurename}{\textbf{Fig.}}

The second class consists of persistent shocks confined near the first two irises, where the continuous gas supply from the 100-kPa upstream source maintains a large pressure difference. The most prominent of these shocks develops into a well-defined Mach disk near the first iris (Fig.~\ref{Fig5}c), exhibiting the characteristic structure of an underexpanded supersonic jet~\cite{franquet2015free,muraoka2022onset}, including an incident shock, a central normal Mach disk, a reflected shock, a triple point, and a slip line (Fig.~\ref{Fig5}d). In conventional free underexpanded jets, the Mach-disk position and diameter vary strongly with the jet-to-ambient pressure ratio. In the present cavity, however, the converging wall and restrictive iris geometrically confine the shock structure near the first-cell exit, similar to the confined Mach disks observed in convergent ducts~\cite{gounko2017patterns}. The Mach disk therefore remains spatially anchored while its precise axial position $x_{\mathrm{MD}}$ and diameter $D_{\mathrm{MD}}$ evolve as the downstream pressure field changes.

A striking connection emerges when these Mach-disk dimensions are compared with the position of the macroscopic gas front. Across the simulated leakage-area ratios for which the Mach disk is resolved ($L_{\mathrm{A}}=0.05$--$0.50$), both $x_{\mathrm{MD}}$ and $D_{\mathrm{MD}}$ collapse onto approximately linear relationships with $x_{\mathrm{f}}$ (Fig.~\ref{Fig5}e and f). Thus, although the primary Mach disk remains confined to a small region near the first iris, its subtle displacement and change in diameter encode the much larger-scale progression of the gas through the entire cavity. This establishes a direct coupling between a localized supersonic structure and the global SLV propagation state.


\section*{Discussion}\label{Sec_discussion}
\noindent The traversal-time scaling and Mach-disk correlation offer complementary insights into two critical challenges in protecting superconducting accelerators: determining the time available for isolation, and enabling early detection of a vacuum breach. The traversal-time scaling can be understood from a simple physical argument. Under choked inflow at fixed upstream conditions, the gas supply rate is approximately proportional to the rupture area. If the amount of gas removed by cryocondensation as the contamination front advances through each cell varies only weakly with breach size, the traversal time should, to leading order, scale inversely with the gas supply rate, yielding $t_{\mathrm{trav}}\sim L_{\mathrm{A}}^{-1}$. The simulated exponent of $-1.19$ is close to this prediction, with the modest deviation likely reflecting the nonlinear evolution of cryocondensation and wall heating. Importantly, this scaling relates breach size to the time available for protective action: detection, decision-making, and valve closure must all be completed before the contamination front reaches the protected boundary. The validated model thus provides a means of assessing this response time window in realistic cavity geometries relevant to large-scale superconducting accelerators such as the European XFEL and LCLS-II~\cite{EuropeanXFEL,LCLSII}.

On the other hand, knowing the available response time is of limited use unless the breach can be detected sufficiently early. Conventional surface thermometers respond only when the slowly advancing contamination front reaches their locations. Along kilometer-scale beamlines, widely spaced thermometers may therefore introduce substantial detection delays, while sufficiently dense sensor coverage would be impractical. The confined Mach disk identified here suggests a promising alternative. Its unsteady pressure field could excite elastic waves in the niobium cavity walls that propagate at kilometers per second~\cite{dellaVentura2025spall}, orders of magnitude faster than the contamination front. These waves could reach sparsely distributed piezoelectric sensors long before the gas front arrives, potentially enabling rapid detection and localization of the breach through acoustic signal analysis. More intriguingly, our simulations reveal that the Mach-disk position and diameter correlate strongly with the advancing gas-front position, suggesting that changes in the associated acoustic signatures could also reveal how far the contamination has propagated. This opens a promising avenue for using a sparse acoustic sensor network not only to detect vacuum failures rapidly, but also to track their progression remotely. Future work will couple the predicted shock dynamics to the structural-acoustic response of the cavity and evaluate the feasibility of this approach.

\section*{References}
\bibliographystyle{naturemag}
\bibliography{VacuumBreak_refs}

\section*{Methods}\label{SecVI}
\noindent\textbf{Frost--wall thermal model}\\
\noindent
To determine the frost-surface temperature entering the cryocondensation law in Eq.~(\ref{eq:mc}), we coupled transient energy balances for the deposited solid-nitrogen (SN$_2$) layer and the underlying metal wall. At each cryogenic wall location, $\delta_{\mathrm{SN}}$ denotes the frost thickness, $T_{\mathrm{s}}$ the temperature at the gas--frost interface, and $T_{\mathrm{w}}$ the temperature of the metal beneath the frost. The mean frost temperature was approximated as $T_{\mathrm{c}}=(T_{\mathrm{s}}+T_{\mathrm{w}})/2$. Using the frost-growth relation given in Results, the local frost energy balance is
\begin{equation}
\rho_{\mathrm{SN}}C_{\mathrm{SN}}\delta_{\mathrm{SN}}
\frac{\partial T_{\mathrm{c}}}{\partial t}
=q_{\mathrm{flux}}-q_{\mathrm{con}},
\label{eq:M_frost_energy}
\end{equation}
where $\rho_{\mathrm{SN}}$ and $C_{\mathrm{SN}}$ are the frost density and specific heat, $q_{\mathrm{flux}}$ is the heat flux delivered by the gas to the frost, and $q_{\mathrm{con}}$ is the heat conducted from the frost into the metal. These fluxes are evaluated as
\begin{align}
q_{\mathrm{flux}}&=\dot m_{\mathrm{c}}
\left(\frac{|\mathbf{u}|^2}{2}+h_{\mathrm{g}}-h_{\mathrm{s}}\right)
+h_{\mathrm{Conv}}(T_{\mathrm{g}}-T_{\mathrm{s}}),
\label{eq:M_frost_input}\\
q_{\mathrm{con}}&=\frac{k_{\mathrm{SN}}}{\delta_{\mathrm{SN}}}
(T_{\mathrm{s}}-T_{\mathrm{w}}),
\label{eq:M_frost_conduction}
\end{align}
where $h_{\mathrm{g}}$ and $h_{\mathrm{s}}$ are the specific enthalpies of gaseous and solid nitrogen, respectively, $h_{\mathrm{Conv}}$ is the convective heat transfer coefficient, and $k_{\mathrm{SN}}$ is the frost thermal conductivity. Thus, heat enters the frost through the kinetic and enthalpy contributions of the deposited gas and through gas-side convection, and leaves by conduction into the wall. The convective heat transfer coefficient is calculated as $h_{\mathrm{Conv}}=k_{\mathrm{g}}Nu/D_i$, where $Nu$ is the gas-side Nusselt number~\cite{baoHeatMassTransfer2020}, $k_{\mathrm{g}}$ the gas thermal conductivity, and $D_i$ the local inner diameter of the corresponding wall element. For the cylindrical tube, $D_i$ is constant, whereas for the cavity, it varies with the axial position $x$ according to the cavity geometry.

The heat conducted through the frost layer raises the temperature of the underlying metal wall, while heat is simultaneously transferred from its outer surface to the surrounding LHe bath. Building on the wall thermal model developed in our previous studies~\cite{baoHeatMassTransfer2020,garceauHeatMassTransfer2021},
we express the local energy balance in a generalized form to accommodate the varying wall geometry of the multicell cavity:
\begin{equation}
\rho_{\mathrm{w}}C_{\mathrm{w}}V_{\mathrm{w}}
\frac{\partial T_{\mathrm{w}}}{\partial t}
=A_{\mathrm{i}}q_{\mathrm{con}}
-A_{\mathrm{o}}q_{\mathrm{He}},
\label{eq:M_wall_energy}
\end{equation}
where $\rho_{\mathrm{w}}$ and $C_{\mathrm{w}}$ are the density and specific heat of the wall material, respectively, and $q_{\mathrm{He}}$ is the heat flux transferred from the outer wall surface to the LHe bath. Here $V_{\mathrm{w}}$ denotes the volume of a local metal-wall element, while $A_{\mathrm{i}}$ and $A_{\mathrm{o}}$ are its inner surface area exposed to the condensing gas and outer surface area in contact with LHe, respectively. These quantities account for the local wall geometry, including the curved surfaces and nearly vertical end walls of the cavity. The wall temperature $T_{\mathrm{w}}$ is assumed to be uniform across the wall thickness, and heat conduction along
the wall surface is neglected. For a straight cylindrical tube with inner diameter $D_i$ and outer diameter $D_o$, the geometric ratios reduce to $V_{\mathrm{w}}/A_{\mathrm{i}}=(D_o^2-D_i^2)/4D_i$ and $A_{\mathrm{o}}/A_{\mathrm{i}}=D_o/D_i$, recovering the cylindrical-wall energy balance used in our earlier model~\cite{baoHeatMassTransfer2020}. Unlike this cylindrical formulation, the generalized balance allows the thermal response of different cavity surfaces to be evaluated using their corresponding local wall geometry.

Temperature-dependent properties of solid nitrogen were obtained from Refs.~\cite{PhysRevB.50.543,konstantinov2005isochoric}, those of nitrogen gas, copper, and stainless steel from Refs.~\cite{stephan_viscosity_1987,flynn_cryogenic_2004}, and those of niobium from Refs.~\cite{koechlin1996parametrization,arblaster2017thermodynamic}.
Frost properties were evaluated at $T_{\mathrm{c}}$, while wall properties were evaluated at $T_{\mathrm{w}}$.

\vspace{12pt}
\noindent\textbf{Heat transfer to liquid helium}\\
\noindent
The heat flux $q_{\mathrm{He}}$ from the metal wall to the surrounding liquid helium was evaluated following the heat-transfer model developed in our previous study~\cite{baoHeatMassTransfer2020}. Since thermal relaxation in liquid helium occurs much faster than gas propagation, quasi-steady heat-transfer correlations were adopted~\cite{van_sciver_helium_2012,smith-1969_Cryogenicsa}. The heat-transfer regime depends on the wall superheat $\Delta T_{\mathrm{w}}=T_{\mathrm{w}}-T_{\mathrm{b}}$, where $T_{\mathrm{b}}$ is the helium bath temperature. For $\Delta T_{\mathrm{w}}<0.1$~K, natural convection was assumed, with $q_{\mathrm{He}}=h_{\mathrm{He}}\Delta T_{\mathrm{w}}$ and $h_{\mathrm{He}}=375$~W/m$^{2}$K~\cite{lantz_heat_2007}. At higher superheat, nucleate boiling occurs and was described by the Kutateladze correlation~\cite{van_sciver_helium_2012}. Beyond a critical heat flux of $q^*=7.5$~kW/m$^{2}$ as determined using the Zuber correlation~\cite{zuber_hydrodynamic_1961}, film boiling occurs and was modeled using the Breen--Westwater correlation $q_{\mathrm{He}}=B_{\mathrm{w}}\Delta T_{\mathrm{w}}^{5/4}$~\cite{breen_effect_1962}, with $B_{\mathrm{w}}=130$~W/m$^{2}$K$^{-5/4}$ following Smith~\cite{smith-1969_Cryogenicsa}.

\vspace{12pt}
\noindent\textbf{Turbulence closure and transport equations}\\
\noindent
To complete the gas-flow formulation presented in Results, the turbulent kinetic energy $k$ and its dissipation rate $\varepsilon$ were modeled using the standard $k$--$\varepsilon$ turbulence model~\cite{Launder1972,Launder1974}, with the following transport equations:
\begin{equation}
\begin{aligned}
\frac{\partial(\rho_{\mathrm{g}}k)}{\partial t}+\nabla\cdot(\rho_{\mathrm{g}}\mathbf{u}k)&=\nabla\cdot[(\mu+\mu_t/\sigma_k)\nabla k]\\
&\quad+P_k-\rho_{\mathrm{g}}\varepsilon+S_k.
\end{aligned}
\label{eq:M_tke}
\end{equation}
\begin{equation}
\begin{aligned}
\frac{\partial(\rho_{\mathrm{g}}\varepsilon)}{\partial t}+&\nabla\cdot(\rho_{\mathrm{g}}\mathbf{u}\varepsilon)=\nabla\cdot[(\mu+\mu_t/\sigma_\varepsilon)\nabla\varepsilon]\\
&\quad+\frac{\varepsilon}{k}(C_{1\varepsilon}P_k-C_{2\varepsilon}\rho_{\mathrm{g}}\varepsilon)+S_\varepsilon.
\end{aligned}
\label{eq:M_dissipation}
\end{equation}
Here $P_k$ represents the production of turbulent kinetic energy by mean-velocity gradients. For axisymmetric flow without swirl, it is expressed as
\begin{equation}
\begin{aligned}
P_k=\mu_t\{&2[(\partial_r u_r)^2+(u_r/r)^2+(\partial_x u_x)^2]\\
&+(\partial_x u_r+\partial_r u_x)^2\},
\end{aligned}
\label{eq:M_production}
\end{equation}
where $u_x$ and $u_r$ denote the axial and radial velocity components, respectively. The sink terms $S_k$ and $S_\varepsilon$ are defined together with the other condensation-induced sinks in Results. We used the standard turbulence-model constants $\sigma_k=1.0$, $\sigma_\varepsilon=1.3$, $C_{1\varepsilon}=1.44$, $C_{2\varepsilon}=1.92$, and turbulent Prandtl number $Pr_t=0.85$. These values, together with $C_\mu=0.09$, are also listed in Supplementary Table~S1.

\vspace{12pt}
\noindent\textbf{Computational domains and boundary conditions}\\
\noindent
All simulations were performed in two-dimensional axisymmetric $x$-$r$ domains. For the experimental validation, the uniform-tube and tube--cell configurations were represented by equivalent axisymmetric geometries that preserved their streamwise lengths and cross-sectional profiles while neglecting the mild curvature of the helical winding. The tube--cell computational domain included the inlet venturi, the upstream stainless-steel transition section, and the LHe-immersed copper test section (Fig.~\ref{Fig2}b). Detailed dimensions and boundary conditions for the individual segments, as well as the time-dependent inlet mass flux reconstructed from the measured nitrogen-reservoir pressure decay, are provided in Supplementary Table~S1. For the nine-cell SRF cavity simulations, the computational domain followed the TESLA cavity profile and the thermal boundary segmentation shown in Fig.~\ref{Fig3}a~\cite{aune2000superconducting}. The wall thicknesses were 2.5~mm for our experimental copper tubes and 2.83~mm for the niobium cavity.

Initially, the cryogenic walls and surrounding LHe bath were at 4.2~K, while the upstream wall sections were assigned thermal conditions appropriate to each configuration. No-slip conditions were imposed at all solid boundaries, with the gas-side temperature of the cryogenic surfaces determined by the evolving frost-surface temperature $T_{\mathrm{s}}$. Along the symmetry axis ($r=0$), the radial velocity vanished ($u_r=0$), together with the radial gradients of axial velocity, pressure, temperature, $k$, and $\varepsilon$. At the downstream boundary, a pressure-outlet condition was imposed, with zero axial gradients of velocity, temperature, $k$, and $\varepsilon$. The inlet conditions differed between the experimental validation and the nine-cell cavity simulations. For the validation cases, the inlet pressure, temperature, and velocity varied with time according to the measured discharge of the nitrogen reservoir~\cite{garceauDesignTestingLiquid2019,baoHeatMassTransfer2020}. For the nine-cell cavity, the incoming flow was assumed to be choked at the vacuum breach ($\mathrm{Ma}_{\mathrm{in}}=1$), with stagnation pressure $P_0=100$~kPa and temperature $T_0=300$~K, as described in Results.

\vspace{12pt}
\noindent\textbf{Numerical solution and convergence}\\
\noindent
The gas equations were solved in ANSYS Fluent with a pressure-based coupled solver. User-defined functions linked the flow calculation to the local condensation and frost--wall thermal models. Within each physical time step, gas pressure, temperature, and velocity in the wall-adjacent cells were passed to the condensation model. The resulting mass-deposition and heat fluxes were used to update the frost and wall temperatures; the revised frost-surface temperature and deposition sinks were then returned to the gas solver. This outer coupling was iterated typically about 20 times per time step (Supplementary Fig.~S3a). Because the condensation sink was confined to the first wall-adjacent cell layer, the induced radial velocity arose from the evolving flow field rather than being imposed as a separate boundary condition (Supplementary Fig.~S3b). The frost was represented as a subgrid layer: its computed thickness remained smaller than the first fluid-cell height, so its growth did not alter the computational mesh.

The frost and wall energy balances, Eqs.~(\ref{eq:M_frost_energy}) and (\ref{eq:M_wall_energy}), were integrated using first-order backward Euler time stepping. For a time step $\Delta t$, we define $A=\rho_{\mathrm{SN}}C_{\mathrm{SN}}\delta_{\mathrm{SN}}/(2\Delta t)$, $B=k_{\mathrm{SN}}/\delta_{\mathrm{SN}}$, and $W=\rho_\mathrm{w} C_\mathrm{w} V_\mathrm{w}/(A_i\Delta t)$. Although the frost thickness was initially zero, a minimum value of $\delta_{\mathrm{SN}}=10^{-9}$~m was used to avoid any numerical singularities. With this treatment, the updated frost-surface and wall temperatures satisfy:
\begin{equation}
\begin{bmatrix}
A+B&A-B\\
-B&W+B
\end{bmatrix}
\begin{bmatrix}
T_{\mathrm{s}}^{n+1}\\T_{\mathrm{w}}^{n+1}
\end{bmatrix}
=
\begin{bmatrix}
2AT_{\mathrm{c}}^n+q_{\mathrm{flux}}^{n+1}\\
WT_{\mathrm{w}}^n-\beta q_{\mathrm{He}}^{n+1}
\end{bmatrix},
\label{eq:M_implicit_thermal_system}
\end{equation}
where $\beta=A_o/A_i$ and $n$ indexes physical time steps. Because the thermal properties and heat fluxes depend on $T_{\mathrm{s}}$ and $T_{\mathrm{w}}$, the coupled equations were solved using a fixed-point iteration at each time step. Starting from the temperatures obtained at the previous time step, the thermal coefficients and heat fluxes were updated iteratively until the changes in both $T_{\mathrm{s}}$ and $T_{\mathrm{w}}$ between successive iterations were below $10^{-4}$~K. This implicit approach ensures numerical stability during the initial stages of frost growth, when the conductive coupling $k_{\mathrm{SN}}/\delta_{\mathrm{SN}}$ is particularly strong.

The validation domains used structured quadrilateral meshes; the nine-cell cavity used a mixture of quadrilateral and triangular elements. In the nine-cell grid study, the coarse, adopted, and fine meshes contained 71,849, 90,238, and 158,172 cells, respectively. Supplementary Fig.~S4 compares the integrated mass-flow histories at the third and ninth irises; the adopted and fine meshes gave closely agreeing results. A typical physical time step was $\Delta t=10^{-6}$~s, and the scaled flow-solver residuals were reduced below $10^{-6}$ except during the initial breach transient.

\vspace{12pt}
\noindent\textbf{Extraction of gas-front and shock characteristics}\\
\noindent
To quantify gas-front propagation from the simulated flow fields, the cross-sectionally integrated axial mass-flow rate $\dot{m}(x,t)$ was evaluated at 1-mm intervals along the cavity, as described in Results. The integration accounts for both forward and reverse axial transport, with recirculating flow contributing negatively to the net mass-flow rate. At each time, the macroscopic gas-front position $x_{\mathrm{f}}(t)$ was determined from the axial distribution $\dot{m}(x,t)$ using a threshold of $0.001$~kg~s$^{-1}$. Because the mass-flow rate can vary nonmonotonically along the cavity, particularly near the irises, multiple axial locations may satisfy this threshold. In such cases, the downstream-most threshold crossing was selected to define $x_{\mathrm{f}}$. The resulting front trajectories were used to determine the cavity traversal times presented in Fig.~\ref{Fig4}.

For shock characterization, the simulated density-gradient magnitude $|\nabla\rho_{\mathrm{g}}|$ and local Mach number $\mathrm{Ma}=|\mathbf{u}|/c_{\mathrm{g}}$, where $c_{\mathrm{g}}$ is the local sound speed, were used to identify shock structures. The primary Mach disk near the first iris was identified by its approximately normal shock front and the associated supersonic-to-subsonic transition along the cavity axis. Its axial position $x_{\mathrm{MD}}$ was determined from the intersection of the normal shock with the symmetry axis, while its diameter $D_{\mathrm{MD}}$ was calculated as twice the radial distance from the axis to the shock triple point (Fig.~\ref{Fig5}d). The extracted values of $x_{\mathrm{MD}}$ and $D_{\mathrm{MD}}$ were compared with the corresponding gas-front position $x_{\mathrm{f}}$ at the same simulation times to obtain the relationships presented in Fig.~\ref{Fig5}e and f.

\section*{Data availability}
\noindent The experimental and simulation data supporting the findings of this study are available from the corresponding author upon reasonable request.

\section*{Code availability}
\noindent All computer codes used in this study are available from the corresponding author upon reasonable request.

\section*{Acknowledgments}
The authors acknowledge the support from the U.S. Department of Energy under Grant No. DE-SC0020113. The work was conducted at the National High Magnetic Field Laboratory at Florida State University, which is supported by the National Science Foundation Cooperative Agreement No. DMR-2128556 and the state of Florida.

\section*{Author contributions}
\noindent W.G. conceived and supervised the research; Y.Q. and Q.R. conducted the experiments and performed the numerical simulations. W.G. and Y.Q drafted the manuscript; all authors discussed the results and reviewed the final manuscript.\\

\section*{Competing interests}
\noindent The authors declare no competing interests.

\end{document}